\documentstyle[12pt,twoside,fleqn,espcrc1]{article}

\newcommand{\AmS}{{\protect\the\textfont2
  A\kern-.1667em\lower.5ex\hbox{M}\kern-.125emS}}

\title{Highlight of the Parallel Session on QCD}

\author{Xin-Nian Wang\address{Nuclear Science Division, MS 70A-3307, 
        Lawrence Berkeley National Laboratory, \\ 
        Berkeley, California 94720}}

\begin{document}
\maketitle

\begin{abstract}
I will give a review with some comments of the parallel session on QCD and the 
related talks in the Quark Matter'97 meeting.
\end{abstract}

\section{Introduction}

As pointed out by T. Matsui in his overview talk \cite{matsui}, QCD is 
really at the heart of the field of ultrarelativistic heavy-ion 
collisions. Virtually every talk in this conference is in some way
related to the study of QCD. It is neither possible nor my 
assignment to summarize the whole conference in just twenty minutes.
I will only give some highlights of one parallel session titled QCD. Even
this is already formidable given the time limitation. Since it is 
unnecessary to repeat what the individual speakers have written in their
contributions to this proceedings, I will give a critical review of their
talks with my own emphasis.

	Though ultrarelativistic heavy-ion collisions are engineered to
discover and study the deconfined phase of nuclear matter, {\it i.e.},
Quark Gluon Plasma (QGP), it encompasses virtually every aspect of QCD
theory, from perturbative QCD (pQCD) hard processes to nonperturbative
hadronization, from parton equilibration to medium modification of hadron
properties. They provide an unprecedented opportunity to study the QCD
theory at vastly different environments. These different aspects of QCD,
dominant at different stages of the high-energy heavy-ion collisions, can
be characterized by the time or energy scales as shown in 
Table \ref{tab:stage}. During the earliest stage of the collisions at
energy scale larger than ,{\it i.e.}, $Q_0=2$ GeV, pQCD processes dominate
and are responsible for production of Drell-Yan dileptons, direct photons,
jets and $J/\Psi$. Minijet production at this stage is also important 
to form a dense partonic matter \cite{minijet}. Among all proposed probes 
of high-energy heavy-ion collisions, these are the only ones whose initial 
production rate can be calculated within the pQCD framework which has been 
successfully tested in $e^+e^-$ annihilation, deeply inelastic $e^-p$ and 
$pp$ or $p\bar{p}$ collision processes \cite{HPC}. The few uncertainties 
involved are the effects of initial state interactions, {\it e.g.}, parton 
shadowing and the so-called Cronin effect, which are also interesting by 
themselves. Talks by Sarcevic~\cite{sarcevic} and Guo~\cite{guo} are 
devoted to these topics. 

At late times when the system is
still dominated by partonic degrees of freedom, interactions among the 
produced partons will drive the system toward equilibrium. The question of
equilibration of the system bears significant importance in the study of
the formation of quark-gluon plasma. It determines how strongly partons
interact with each other in the system and how long the partonic phase
will last. During this period of time, or the life-time $t_{\rm QGP}$ of 
the parton system, there will be associated thermal production of particles 
which can also be used as signals of the dense QGP matter. Talks by 
Sakai~\cite{sakai} and Wong~\cite{wong} address some issues in this stage.

\begin{table}[hbt]
\caption{Physics at different stages of high-energy heavy-ion collisions}
\label{tab:stage}
\begin{tabular}{|c|c|c|} \hline
$t$(fm) & $Q$(GeV) & Physics \\ \hline
$\leq 0.1$ & $\geq 2$ & \parbox[t]{4in}{pQCD: Drell-Yan, $J/\Psi$,
	direct $\gamma$, and jets production} \\ \hline
$t_{QGP}$& $1\sim 2$ & \parbox[t]{4in}{pQCD or Finite temperature QCD:
	Parton equilibration, thermal production of particles, $J/\Psi$
	suppression, and jet quenching} \\ \hline
$t_{Mix}$ & $\Lambda_{QCD}\sim 0.2$ & \parbox[t]{4in}{Hadronization 
	(in equilibrium?): Partition in phase space and flavor which
	may be modeled by ``fireball model''.} \\ \hline
$t_H$ & $< \Lambda_{QCD}$ & \parbox[t]{4in}{Hadronic interaction: medium 
	effects, chiral condensates, etc, studied in effective models.}
	\\ \hline 
\end{tabular}
\end{table}

When the energy scale in the system drops to around the value of 
$\Lambda_{\rm QCD}$ hadronization will convert partons into hadrons.
The process is purely nonperturbative and so far there has not been
any known description of this process from QCD theory. It is not clear
at all whether the system is in equilibrium during the hadronization 
process. When the hadronization happens in the vacuum as in $e^+e^-$, $ep$,
$pp$ and $p\bar{p}$ collisions, phenomenological parameterizations of
the so-called fragmentation functions are normally used. An exponential
form motivated by vacuum tunneling in a strong field is normally used to 
describe the mass and transverse momentum distribution of the produced 
particles~\cite{lund}. Under different environments such as in high-energy
heavy-ion collisions, the parameters governing the hadronization will also
be different. If the hadronization happens in a nonequilibrium environment,
then the signals of the initial strangeness
 enhancement during the partonic phase
might easily get lost in the hadronization process which also produces 
strange particles. One might also use statistic approach as by Becattini 
{\it et al} \cite{f_becattini} to describe the particle production based 
on the occupation of the phase space. However, one must bear in mind that
the so-call temperature parameter extracted from such an analysis has nothing
to do with the temperature of an equilibrated system and thus cannot be put
on the usual phase diagram.

After the hadronization stage, the interaction among hadrons might be
described by some effective theories in which one can discuss physics
phenomena such as medium effects in hadron properties and disoriented 
chiral condensates. The talk by Mishustin~\cite{mishustin} discusses
a model of chiral phase transition. Talks by Alam, Rajagopal and 
Sch\"afer~\cite{alam,shafer,rajagopal} demonstrate the renewed interests in the
physics at high baryon densities. In the following, I will summarize these
talks in the QCD parallel session with some comments.

\section{Effects of Initial-state Parton Interactions}

All of the hard probes of the quark-gluon plasma involve hard parton 
scattering which can be studied within the framework of perturbative QCD.
The current analysis of the $J/\Psi$ data from NA50 \cite{jpsi} has already
demonstrated how important it is to understand accurately the initial
production rates of these hard probes and their spectra. One important 
problem one has to study in this aspect is the effect of initial-state
parton scatterings. Only after such effects have been completely understood,
can one disentangle the true QGP signals from other conventional nuclear 
effects. Initial parton interactions and the interference in them can lead
to an apparent depletion of the effective parton density inside a nucleus,
the so-called nuclear shadowing of the parton distributions. They can also
modify the momentum spectra of the produced hard probes which is often
referred to as Cronin effect \cite{cronin}.

	There are many models for the nuclear shadowing of the parton 
distributions. A partonic picture with Glauber-Gribov interference effect
along the line of a similar study by Levin {\it et al.} \cite{levin} was 
reported by Sarcevic. In this model the virtual photon in deeply inelastic 
$e^-p$ (or $e^-A$) scatterings is converted into a $q\bar{q}$ pair first which
then interacts with the proton (or nucleus) via Pomeron exchanges. The proton
(or nucleus) structure function $F_2^{p(A)}$ can then be directly related to
the cross section of this $q\bar{q}$ pair with the proton ( nucleus).  One can
then derive the DGLAP evolution equation for the parton distributions.
They assume that eikonalization can be applied to the coherent multiple 
scatterings of the $q\bar{q}$ pair with the nucleus (at small $x$ the 
coherent length of the $q\bar{q}$ pair is much larger than the nuclear 
size, $1/m_Nx\gg R_A$), so that

\begin{eqnarray}
\sigma_{q\bar{q}A}&=&2 \int d^2b [1-e^{-\sigma_{q\bar{q}N} T(b)/2}] \nonumber\\
	&=&A\sigma_{q\bar{q}N} \left( 1 - 
	\frac{A\sigma_{q\bar{q}N}}{8\pi R_A^2} +\cdots\right).
\end{eqnarray}

Since $\sigma_{q\bar{q}N}$ and $\sigma_{q\bar{q}A}$ are related to the 
structure function of a nucleon and nucleus respectively, one can then 
derive the DGLAP evolution equation for the nucleus structure function
which depends on the gluon distribution in a nucleon. It is interesting
to point out a term in the evolution equation arising from the second term
in the above equation that corresponds exactly to the gluon fusion term in the
Mueller and Qiu's \cite{qiu} model of shadowing. With some choice
of the initial parton distributions in nuclei ( at a fixed scale $Q_0$), they
reproduced the measured nuclear shadowing effect of the structure function.
They further predicted the shadowing effect for gluon distribution using the
approach and concluded that the depletion of gluon density will not saturate
at very small $x$. I believe this is only because they did not take into
account similar shadowing due to gluon fusion in a nucleon. At very small
$x$ the same mechanism causes the gluon distribution in both a nucleon and
a nucleus to saturate and then the depletion inside a nucleus due to shadowing
will also saturate.

Initial parton scatterings not only cause the depletion of the effective parton
distribution inside a nucleus, it can also modify the momentum spectra of the
produced particles in the hard processes, like the $p_T$ distribution of 
DY lepton pairs, direct photons and $J/\Psi$. There has been a 
continuing effect
by Qiu, Sterman and collaborators \cite{qiust} to study such effects within
the context of double scatterings which corresponds to the next-leading-twist
contribution to the hard processes. Guo \cite{guo} reported her recent 
calculation along the same line of double scatterings in DY dilepton 
production in $pA$ collisions. One can separate the contribution into three
terms depending on the momentum involved in the second 
scattering in addition to the $q\bar{q}$ annihilation. If the  
the second scattering is hard, the contribution has a form of 
classical double scattering, $\sigma_1\sigma_2/Q^2$. If the the second
is soft, the contribution is proportional only to $\sigma_1$ with the
coefficient depending on a universal twist-4 nuclear parton correlation 
function which can be determined from other processes. The third term
is then the interference between the first two amplitudes. For large
transverse momentum $q_T$ of the lepton pair which Guo has considered, the 
interference term is small. The double scattering will then enhance the 
total cross section.  When $q_T$ is small, the interference term is 
expect to be large and nagative so the contribution from the double
scattering will be reduced. But for small $q_T$, one has to take into
account of the resummation of soft gluon radiation. This is still being
investigated.

As we see from both Gao and Sarcevic's talks and as again emphasized in 
Dokshitzer's review talk \cite{dok}, quantum interference effects are 
important in many QCD processes including high-energy heavy-ion 
collisions. These important effects have to be considered when one
tries to construct a reasonable model. Any model, no matter how open it
is, is useless if it does not have the right physics in it. With
the coming RHIC experiments and  also the high $p_T$ region the SPS
experiments have so far reached, the hard processes will become a test
ground for many QCD phenomena and powerful probes of the dense matter.

\section{Equilibration Processes in a Parton Gas}

While the initial production of the hard probes and minijets can be 
studied in the framework of pQCD, the early evolution of the produced
parton gas can also be similarly addressed via pQCD based parton cascade
approach. In the talk by Wong \cite{wong}, several improvements are reported
on the evolution of the initially produced parton gas based upon a
Boltzman approach \cite{wang}. He included not only the elastic but also
inelastic processes of parton-parton scattering in the calculation of the 
relaxation time in the relaxation-time approximation of the Boltzman equation.
They considered radiative correction to the problem and find it important
when the strong coupling constant is not very small. It then leads to smaller
relaxation time and thus faster equilibration. On the other hand it also
accelerates the cooling of the temperature and reduces the life-time of
the system (the time before the hadronization happens when temperature
drops below the QCD phase transition temperature). When one calculates the
yields of thermal particle production during the life-time of the partonic
system, these two effects will then compete with each other. The final results
will also be very sensitive to the initial conditions one uses and that is the
main uncertainty one has to deal with in the partonic approach of heavy-ion
collisions.

	In a related effort, calculation of the transport coefficients was
performed within a lattice QCD approach \cite{sakai}. What are really 
calculated in this case are the the Matsubara Green's functions at finite
temperature and by analytic continuation the retarded Green's functions.
The transport coefficients are then obtained in terms of the retarded Green's
functions in the linear response theory. The coefficients are found at around
the QCD critical temperature very close to the perturbative calculation at
high temperatures. One, however, should be reminded that in this approach
one has to assume an ansatz for the form of the spectrum density of the
Fourier transform of the Green's functions.

\section{Physics at High Baryon Density}

During the last year, there has been a renewed interest in the physics of
dense matter, in particular the possibility of diquark condensates and
the associated color superconductivity. The idea of color superconductivity
in the matter of high baryon density proposed a long time ago \cite{bailin}
is quite simple. The possible states of a pair of quarks, which are 
color-triplet and spin-1/2 objects, are color anti-triplet (antisymmetric) 
and sextuplet (symmetric) and spin zero and one, {\it i.e.}, with color
and spin quantum numbers $(\bar{3},0)$, $(\bar{3},1)$, (6,0), and (6,1). Out
of these six possible diquark states, the most attractive is the color-
anti-triplet and spin-zero combination $(\bar{3},0)$. If the quarks are
in a spatial symmetric configuration in this state, the Fermi statistics
requires the state to have two quarks with different flavors. This is
consistent with the diquark structure of a baryon where color confinement
also requires the diquark to be in an color-anti-triplet state. Such a diquark
description of a baryon has been successfully implemented in the Lund model 
of string fragmentation \cite{lund} and is also consistent with the results
of a QCD sum rule study of the quark distribution inside a baryon \cite{cz}. 
Therefore, in a baryon dense (and chirally symmetric) system  quarks tend 
to form iso-singlet spin-zero and color-anti-triplet pairs, if color 
confinement is not a necessary condition. This could then happen in a system 
with very high baryon density where the color could well be deconfined. 
At low enough temperature the boson-like quark pairs will then go through 
Bose-Einstein condensation and give rise to nonvanishing diquark 
condensates $<qq>$.  This system will then resemble a cool electron plasma
and the quark pairs bounded together by the attractive Coulomb interaction
will be like Cooper pairs on the Fermi surface, thus leading to color
superconductivity. Using a perturbative calculation, the corresponding
gap and critical temperature were estimated \cite{bailin} to be in the 
MeV range.

	Recent two studies \cite{shafer,rajagopal} presented in this 
conference argue that non-perturbative interaction among quarks can also 
lead to the same phenomenon but with the gap and critical temperature two
orders of magnitude larger. Both studies have used an instanton-induced
interactions between light quarks. They also found that the iso-singlet
scalar diquark with color-anti-triplet is the most attractive channel.
The corresponding gap and critical temperature are found in the range of
100 MeV. 

	It is interesting to point out that the formation of diquark 
condensates requires deconfinement at high baryon density. But 
the instanton-like interaction could also be screened in such a
deconfined phase so that the dominant interaction again will be reduced
to Coulomb-like interaction with one gluon exchange. In Bailin and Love's
calculation, they used an effective strong coupling constant of
$\alpha_s=0.3$.  Iwasaki and Iwado \cite{iwado}
 recently pointed out that if one
uses a larger value of the effective $\alpha_s$, one can also get larger
values of the gap and critical temperature. Therefore, no matter what
type of interaction is dominant, color superconductivity is still
an interesting possibility. Since the diquark condensates has to be in one
particular direction in the color space, it will also force the color
of the quarks to be aligned in the opposite direction in order to maintain 
the color neutrality of the system. Then the question is what are the
consequences that one observe, for example in neutron stars.

This work was supported by the Director, Office of Energy Research, Division
of Nuclear Physics of the Office of High Energy and Nuclear Physics of the
US Department of Energy under Contract No. DE-AC03-76SF00098.

\end{document}